\documentclass[10pt, twocolumn, pre, superscriptaddress,floatfix] {revtex4}
\usepackage{graphicx, calc, bm, epsfig}
\newcommand{\toff}{~$t_\mathrm{off}$}
\newcommand{\ton}{~$t_\mathrm{on}$}

\begin{document}

\title{Mechanical coupling in flashing ratchets}

\author{Erin M. Craig}
\affiliation{Materials Science Institute and Physics Department, University of
  Oregon, Eugene, Oregon 97405, USA}
\author{Martin J. Zuckermann}
\affiliation{Department of Physics, Simon Fraser University, Burnaby, B.C. V5A
  1S6, Canada}
\author{Heiner Linke}
\affiliation{Materials Science Institute and Physics Department, University of
  Oregon, Eugene, Oregon 97405, USA}
\date{\today}

\begin{abstract}
We consider the transport of rigid objects with internal structure in a flashing ratchet potential by investigating the overdamped behavior of a rod-like chain of evenly spaced point particles. In 1D, analytical arguments show that the velocity can reverse direction multiple times in response to changing the size of the chain or the temperature of the heat bath. The physical reason is that the effective potential experienced by the mechanically coupled objects can have a different symmetry than that of individual objects. All analytical predictions are confirmed by Brownian dynamics simulations. These results may provide a route to simple, coarse-grained models of molecular motor transport that incorporate an object's size and rotational degrees of freedom into the mechanism of transport.
\end{abstract}
\maketitle

\section{Introduction\label{introduction}}

A number of theoretical and experimental studies in recent years have addressed directed transport of diffusive particles in spatially periodic systems in the absence of net forces, which occurs when spatial or temporal inversion symmetry is broken while the system is kept away from thermal equilibrium \cite{hanggi, reimann3}. Such systems are called ratchets or Brownian motors \cite{bartussek}, and their study has both fundamental and practical motivations, including applications in biology and nanotechnology \cite{linke}. There are various mechanisms by which the system can be maintained out of thermal equilibrium, including a time-dependent force (often called a ``rocking" ratchet) \cite{magnasco, bartussek2, kostur2} and a time-dependent potential barrier (a ``flashing" ratchet) \cite{ajdari, bug}.

While many ratchet studies have dealt with the asymmetric pumping of individual, pointlike particles \cite{ajdari, astumian, magnasco, doering}, motors with internal structure have recently attracted interest. One reason is that such studies may provide a route to models of the linear transport of biological molecular motors. A number of studies have demonstrated behavior of mechanically coupled particles which is qualitatively different than that of a single particle \cite{cilla, dan, derenyi2, klumpp, stratopoulos, igarashi1, julicher1, julicher2, julicher3, csahok, ajdari2, denisov, downton, li, aghababaie, derenyi, dialynas2, dialynas1, wang}, for example: In contrast to the behavior of an individual particle, two harmonically coupled particles in a flashing ratchet undergo directed motion in the absence of thermal fluctuations \cite{ajdari2, klumpp}. For a flashing ratchet, two harmonically coupled particles have a smaller velocity than a single particle, while for a rocking ratchet, the coupled particles have a greater velocity \cite{wang}. Two rigidly coupled particles in the presence of a stochastically rocked ratchet potential undergo transport that reverses direction as a function of the dimer size \cite{dialynas2, dialynas1}. Large collections of mechanically coupled particles in a flashing ratchet can undergo spontaneous oscillations, which may be relevant to the collective motion of molecular motors in muscles \cite{julicher1}. 
 
\begin{figure}[htbp]
  \centering
  \includegraphics[width=\columnwidth]{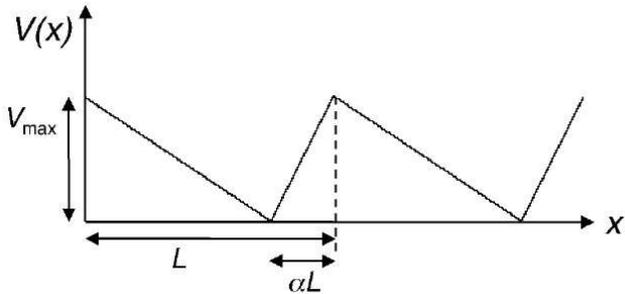}
  \caption{The applied ratchet potential $V(x)$, shown above, is characterized by periodic length, $L$, height, $V_\mathrm{max}$ and asymmetry, $\alpha$.}
  \label{fig:fig1}
\end{figure}

In this study, we wish to establish a conceptual understanding of the role of coupled motion in a flashing ratchet by considering the most simple possible form of mechanical coupling between particles: a chain of rigidly connected point particles. The chains are exposed to Gaussian white noise in the overdamped limit, and driven by a periodically modulated, asymmetric, spatially periodic potential (Fig.~\ref{fig:fig1}). The periodically flashing ratchet scheme is chosen because it is simple enough to allow an intuitive understanding and analytical prediction of the behavior of coupled particles. 

Using analytical arguments, we show that, for a 1D system, the average velocity can reverse direction in response to changing the size of the chain or the temperature of the heat bath without changing the symmetry of the applied potential. However, when the chain is allowed to rotate freely in 3D, or when its length is much less than the spatial period of the ratchet potential, velocity reversal is no longer observed, and the qualitative behavior of single particle motion is recovered. The behavior observed for mechanically coupled particles in a flashing ratchet can be understood in terms of the chain's center-of-mass effective potential, which can have different symmetry than the potential felt by an individual particle. All analytical predictions are confirmed by numerical simulations.

The layout of this paper is as follows: In section \ref{model}, we introduce the Brownian dynamics (BD) model used in our simulations. In section \ref{lowtemp}, we discuss the motion of chains of two or more evenly spaced point particles constrained to 1D motion in the limit of perfect confinement ($kT \ll V_\mathrm{max}$). In section \ref{temp}, we discuss how the direction of ratchet velocity for two coupled particles moving in 1D depends on the temperature of the surrounding heat bath. The behavior of rigidly connected particles in 3D is discussed in the section \ref{3Dmotors}. In the conclusion, we discuss a possible experimental realization of this system, which may be used to directionally separate polymer segments of different length. We also discuss theÊ possible biological relevance of our results, namely the observation that complex molecular motors with only small differences in structure can move in opposite directions.

\section{The model\label{model}}

We consider mechanically coupled particles periodically subjected to a piecewise linear `sawtooth' potential, $V(x)$ (Fig.~\ref{fig:fig1}), characterized by periodic length, $L$, height, $V_\mathrm{max}$, and asymmetry, $\alpha$. The particles are alternatingly subjected to $V(x)$ for a time period \ton, and allowed to diffuse freely for a time period \toff, thus asymmetrically harnessing Brownian motion to produce net transport which, for $\alpha<1/2$, is in the $+x$ direction in Fig.~\ref{fig:fig1} for non-interacting particles (see section \ref{lowtemp} and \cite{reimann3, ajdari, ajdari2}). 

The simulations in this study are based on the following scheme:  The coupled particles are modeled as a chain of point-like beads, separated by a distance $d$, defined by a repulsive Lennard-Jones interaction 

\begin{equation}
\label{eq:lj}
 V_{ij}(r_{ij})=\left\{\begin{array}{r@{\quad:\quad}l}
      4\varepsilon\left((\frac{\sigma}{r_{ij}})^{12} -
      (\frac{\sigma}{r_{ij}})^{6}\right) +\varepsilon & r_{ij}\le
      2^{\frac{1}{6}}\sigma\\
      0 & r_{ij}>2^{\frac{1}{6}}\sigma\\
  \end{array}
  \right. ,
\end{equation}
where $r_{ij}$ is the separation between beads, and a finite extensible nonlinear elastic (FENE) potential between adjacent beads: 

\begin{equation}
\label{eq:fene}
 U(r_{ij})=-\frac{1}{2}k_{\mathrm{F}}R_{0}^2\ln\left(1 - \frac{r_{ij}^2}{R_0^2}\right).
\end{equation}

The equations of motion of individual beads are given by 

\begin{equation}
\label{langevin}
m\ddot{\mathbf{r_i}}=-\gamma_b \dot{\mathbf{r_i}}+\xi(t) - \nabla V_{INT}(\mathbf{r_i}) - \nabla V_{EXT}(t, \mathbf{r_i}) ,
\end{equation}
where $\xi(t)$  is a randomly fluctuating Gaussian white noise term with zero mean and correlation  $\left<\xi\mbox{(t)}\xi\mbox{(s)}\right>=2\gamma_b \mbox{kT}\delta\mbox{(t-s)}$, $\gamma_b$ is the drag coefficient of a bead, $k$ is the Boltzmann constant and $T$ is the temperature of the heat bath. The term $V_{INT}(\mathbf{r_i})$ represents the potential of the bead due to its interaction with the other beads in the chain, and $V_{EXT}(t, \mathbf{r_i})$ is the external ratchet potential. The equations of motion are integrated using a Brownian dynamics algorithm. The mass, $m$, of each bead is set to unity and we use $\sigma$, $\varepsilon$ and $\tau=\sqrt{\frac{m\sigma^2}{\varepsilon}}$  as scaled units of length, energy and time. 

Unless otherwise specified, $L=5 \sigma$, and $R_0$ is chosen such that the distance between adjacent beads is $d=0.97\sigma$ for $kT=\varepsilon$. In sections \ref{dimer} and \ref{trimer}, the FENE parameter $R_0$ is altered to adjust the value of $d$. In section \ref{temp}, the temperature $T$ is varied, but $R_0$ is adjusted accordingly, so that $d$ is the same for different values of the temperature. We set \ton~= \toff~= 20, because this gives a chain of several beads enough time to localize during \ton~and to diffuse to an adjacent well during \toff. 

In order to isolate the role of an object's geometry in ratchet transport from the effect of varying the diffusion constant, we give all chains the same total drag coefficient. The drag coefficient for one bead is chosen to be $\gamma_b=1/N$, where $N$ is the number of beads in the chain. Since hydrodynamic effects have not been included, the chain's total drag coefficient is $\gamma_T=N\gamma_b=1$ \cite{grosberg}. 

When the particles are constrained to motion along the $x$-axis, this model describes a rigid rod made up of evenly spaced particles. When individual particles are allowed to move in three dimensions, this system corresponds to the Rouse model of a polymer, with equilibrium configurations described by a self-avoiding walk \cite{kremer, downton}. This modeling scheme is used here for rigid rods in order to provide continuity with our investigation of the role of flexibility in flashing ratchet transport \cite{downton}.

\section{Effects of coupling in low temperature regime\label{lowtemp}}

Before we describe how mechanically coupling the motion of point particles affects their ratchet velocity, we briefly review the motion of a single particle (a monomer) in a flashing ratchet \cite{reimann3, ajdari, ajdari2}. In the low-temperature limit ($kT \ll V_\mathrm{max}$), a monomer will have positive velocity when $\alpha < 1/2$ for the following reason: During \ton, the monomer localizes at the minimum of the sawtooth potential: $x_\mathrm{min}=(1-\alpha)L$, where $x = 0$ corresponds to the beginning of the potential well. The minimum distance the monomer must diffuse during \toff~in order to localize in the adjacent well in the $+x$ direction during \ton~is $\Delta x_+ = \alpha L$. The minimum diffusion distance to localize in the adjacent well in the $-x$ direction is $\Delta x_- = (1- \alpha )L$. The ratchet velocity will be positive if $\triangle x_{+} < \triangle x_{-}$. As the asymmetry, $\alpha$, is increased from zero, there is a reversal from positive to negative velocity at a critical asymmetry, $\alpha_c = 1/2$, given by the condition: $\triangle x_{+} = \triangle x_{-}$. 

For a 1D chain of particles in the limit $kT \ll V_\mathrm{max}$, the characteristic diffusion distances, $\Delta x_+$ and $\Delta x_-$, can be calculated for the chain's center of mass (CM) by considering the shape of the effective potential, $U(x_\mathrm{CM})$, given by

\begin{equation}
\label{effectivepotential}
U(x_\mathrm{CM})=\frac{1}{N}\sum_{i=1}^NV(x_i).
\end{equation}

The center of mass of a chain of coupled particles will localize exactly at a local minimum, $x_\mathrm{min}$, of its effective potential. The distance $\Delta x_+$ ($\Delta x_-$) is given by the distance between $x_\mathrm{min}$ and the closest absolute effective potential maximum in the $+x$ ($-x$) direction (Fig.~\ref{fig:fig2}). The net ratchet velocity is positive (negative) for $\alpha < \alpha_c$ ($\alpha > \alpha_c$), respectively, where $\alpha_c$ is again given by the condition $\Delta x_+ = \Delta x_-$. For simplicity, we limit our discussion to chains with total length less than a spatial period of the ratchet ($(N-1)d \leq L$). The velocity of any object is antisymmetric about $\alpha=1/2$, so only values of $\alpha$ in the interval $[0; 0.5]$ are discussed in this study.

\subsection{Dimer\label{dimer}}

We begin to explore the role of coupled particle motion in a flashing ratchet by considering two coupled particles (a dimer) separated by a distance $d$, constrained to 1D motion. The system behaves differently depending on whether $d< \alpha L$ or $d > \alpha L$, and we will briefly discuss each of these cases. 

\begin{figure}[htbp]
  \centering
  \includegraphics[width= \columnwidth]{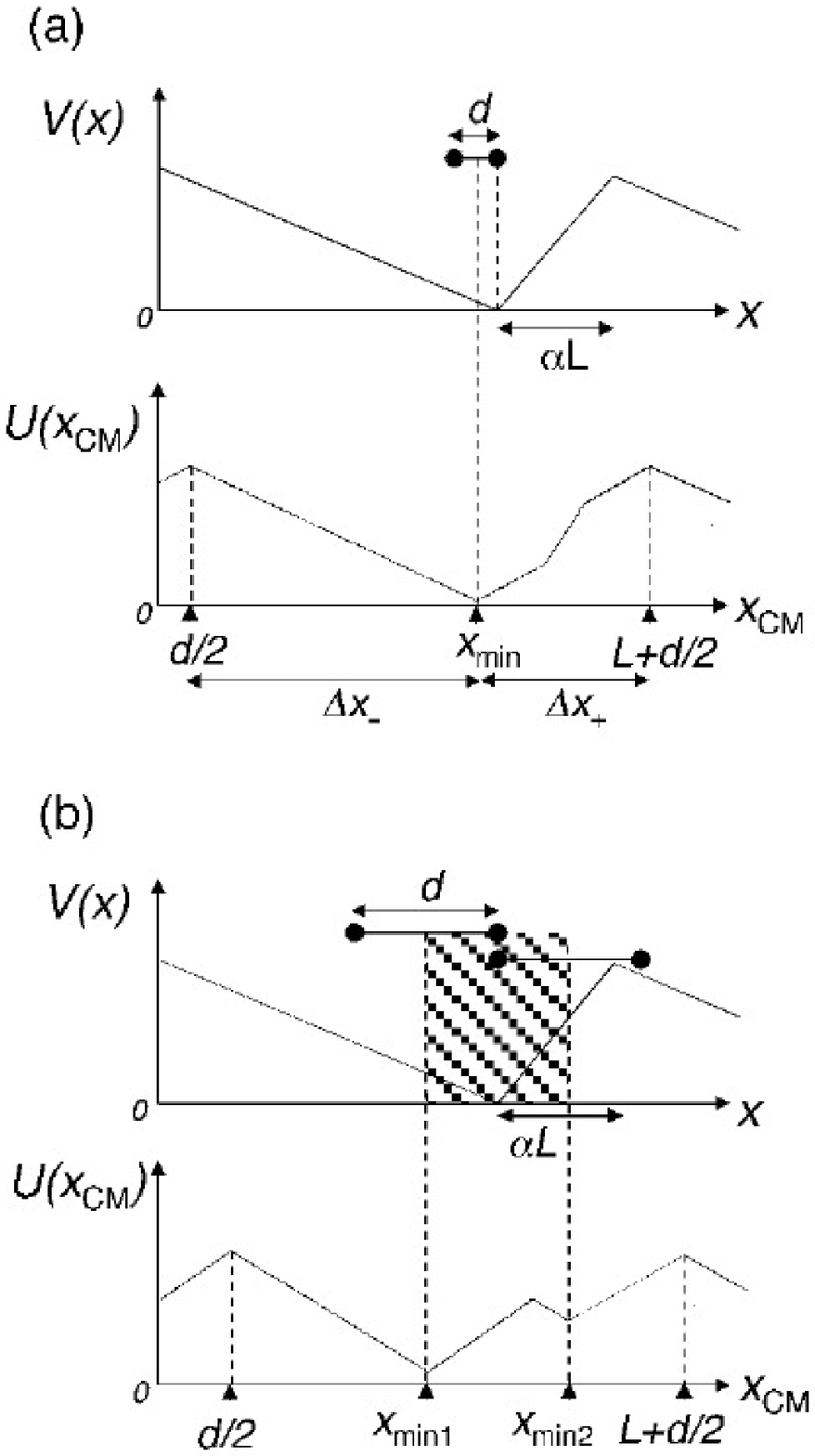}
  \caption{(a) Schematic of 1D ratchet potential $V(x)$ with $\alpha=0.25$ (top) and the effective potential $U(x_\mathrm{CM})$ of a dimer ($d=0.2L$) in this potential. The dimer's localization position is $x_\mathrm{min} = (1-\alpha) L - d/2$. The diffusion distances necessary for localization in the adjacent wells in the $+x$ direction and in the $-x$ direction are labeled as $\Delta x_+$ and $\Delta x_-$ respectively. (b) $V(x)$ and $U(x_\mathrm{CM})$ for $\alpha=0.25$ and a dimer of length $d=0.33 L$. Since now $d > \alpha L$, there are two possible localization positions. The dimer's average localization position falls somewhere in the shaded region (see text).}
  \label{fig:fig2}
\end{figure}

When $d<\alpha L$, the dimer localizes to the center-of-mass position $x_\mathrm{min}=(1-\alpha)L-d/2$ during $t_\mathrm{on}~$  because, at this position, one bead experiences the shallow slope of $V(x)$ and the other sits at the minimum of the potential well. Any displacement in the $+x$ or $-x$ direction would lead to a restoring total force, and so $x_\mathrm{min}$ corresponds to a minimum of $U(x_\mathrm{CM})$ (Fig.~\ref{fig:fig2}(a)).

For a dimer to localize in the adjacent well in the $+x$ direction after a \toff~ period, it must diffuse until $x_\mathrm{CM} > L+d/2$, such that both beads are in the next well and the dimer experiences a total force $F_\mathrm{CM}>0$ when the potential turns on. Therefore, $\triangle x_{+} = L+d/2 - x_\mathrm{min} = \alpha L + d$. 

Likewise, for a dimer to localize in the adjacent well in the $-x$ direction after a \toff~ period, it must diffuse until $x_\mathrm{CM} < d/2$, such that one bead is exposed to the steep side of the adjacent well in the $-x$ direction, and $F_\mathrm{CM}<0$ at the beginning of \ton. Consequently, $\triangle x_{-} = x_\mathrm{min} - d/2 = (1-\alpha)L - d$, and the condition, $\triangle x_{+} = \triangle x_{-}$, yields the critical asymmetry

\begin{equation}
\label{acrit1}
\alpha_c=\frac{1}{2}-\frac{d}{L}.
\end{equation}

Since Eq.~(\ref{acrit1}) was obtained by assuming $d< \alpha L$, it holds for $d/L<\alpha_c=1/2-d/L$, or $d/L<1/4$.

When $d > \alpha L$, the effective potential $U(x_\mathrm{CM})$ has two local minima: $x_{min1}=(1-\alpha) L -d/2$ and $x_{min2}=(1-\alpha) L+ d/2$, where $x_{min1}$ corresponds to localization inside a $V(x)$ potential well, and $x_{min2}$ corresponds to a dimer straddling a $V(x)$ potential maximum (Fig.~\ref{fig:fig2}(b)). The average position during \ton~depends on the relative probability for the object to be trapped in each of these minima, which is a function of $\alpha$, $d/L$ and \toff. If the dimer has a probability $\Phi \left(\alpha,\frac{d}{L},\mbox{\toff} \right)$ of being trapped in the position $x_{min1}$ at the end of \toff, then the average localization position is:  $\left<x_\mathrm{min}\right>=\Phi x_{min1} +(1-\Phi )x_{min2}=(1-\alpha )L+(1-2\Phi )d/2$. 

In this case, the characteristic diffusion distances are $\triangle x_{+} = L+d/2 - \langle x_\mathrm{min} \rangle$ and $\triangle x_{-} = \langle x_\mathrm{min} \rangle - d/2$, and the condition, $ \triangle x_{+} = \triangle x_{-}$, yields

\begin{equation}
\label{acrit2}
\alpha_c=\frac{1}{2}-\Phi \left(\alpha_c,\frac{d}{L},\mbox{\toff} \right)\frac{d}{L}.
\end{equation}
Note that for $d<\alpha L$, where $\Phi = 1$, Eq.~(\ref{acrit2}) simplifies to Eq.~(\ref{acrit1}).

\begin{figure}[htbp]
  \centering
  \includegraphics[width= \columnwidth]{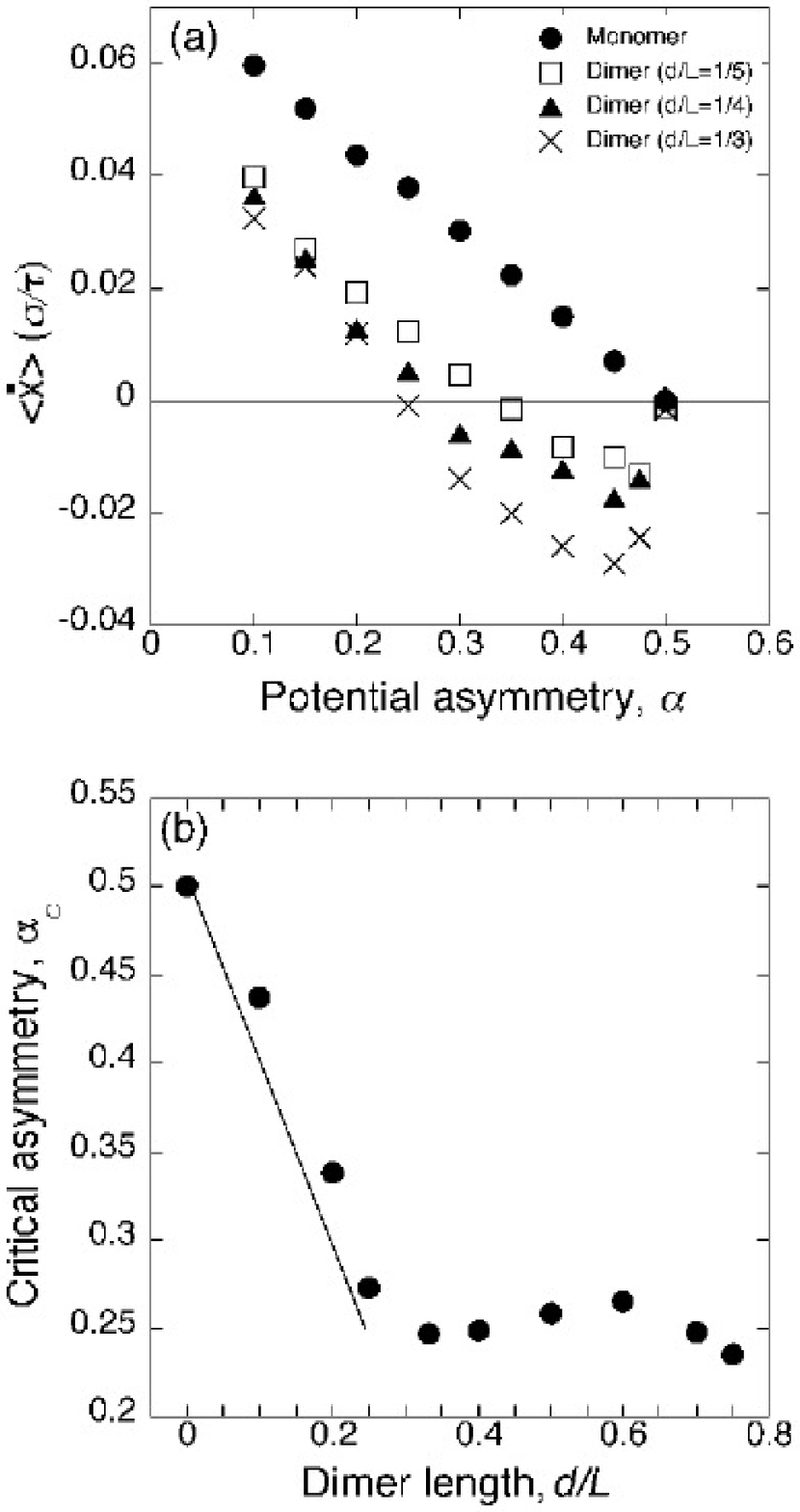}
  \caption{Brownian dynamics simulations. (a) Average velocity as a function of $\alpha$ for a monomer and for dimers of length $d=L/5$, $L/4$ and $L/3$, using $kT/V_\mathrm{max}=1/50$, $L=5.6$ and $\mbox{\ton}=\mbox{\toff}=20$. (b) Critical asymmetry, $\alpha_c$, as a function of $d/L$. Points below $d/L=1/4$ are compared with the prediction of Eq.~(\ref{acrit1}), $\alpha_c=1/2-d/L$ (solid line).}
  \label{fig:fig3}
\end{figure}

These analytical predictions are confirmed by BD simulations. Figure~\ref{fig:fig3}(a) shows simulation results for the time-averaged velocity of dimers with different $d$, for $kT/V_\mathrm{max} = 1/50$, demonstrating that the ratchet velocity reverses as a function of $\alpha$. Figure~\ref{fig:fig3}(b) confirms the prediction of Eq.~(\ref{acrit1}) that $\alpha_c$ decreases linearly with $d/L$, for $d/L<1/4$, and is no longer linear with $d/L$ for $d/L>1/4$ (Eq.~(\ref{acrit2})).

\subsection{Three or more coupled particles\label{trimer}}
 
We now extend our discussion to chains of three or more particles moving in 1D. In particular, we make analytical arguments for when a velocity reversal with $\alpha$ is expected for three coupled particles (a trimer), and support these predictions with BD simulations. Analytical predictions become increasingly complicated for $N>3$, and we simulate chains with $N=4$ and $N=5$ to make general observations about 1D coupled motion in a flashing ratchet.

For $N$ coupled particles, each separated by a distance $d$, the number and locations of minima in the effective potential depend on the following factors: (1) Is the potential asymmetric enough that one particle on the short (steep) side will experience more force than $(N-1)$ particles on the long (shallow) side (i.e. is $\alpha < 1/N$)? If $\alpha<1/N$, the object will localize at a position where none of the particles is on the steep part of the potential. If $\alpha > 1/N$, the force exerted on particles in the shallow part of the potential will push the localization position further to the +x direction. (2) Is the separation between particles longer than the short side of the sawtooth potential (i.e. is $d>\alpha L$)? When $d< \alpha L$, each period of the effective potential $U(x_\mathrm{CM})$ will have only one minimum. For $d > \alpha L$, there are $N$ minima for each period of $U(x_\mathrm{CM})$, each corresponding to one of the particles localizing at the minimum of the sawtooth potential $V(x)$.

Based on these factors, we discuss the behavior of a trimer in the following four regimes: (I) $\alpha<1/N=1/3$ and $d<\alpha L$; (II) $\alpha<1/N=1/3$ and $d>\alpha L$; (III) $\alpha>1/N=1/3$ and $d<\alpha L$; and (IV) $\alpha>1/N=1/3$ and $d>\alpha L$.

\begin{figure}[htbp]
  \centering
  \includegraphics[width=\columnwidth]{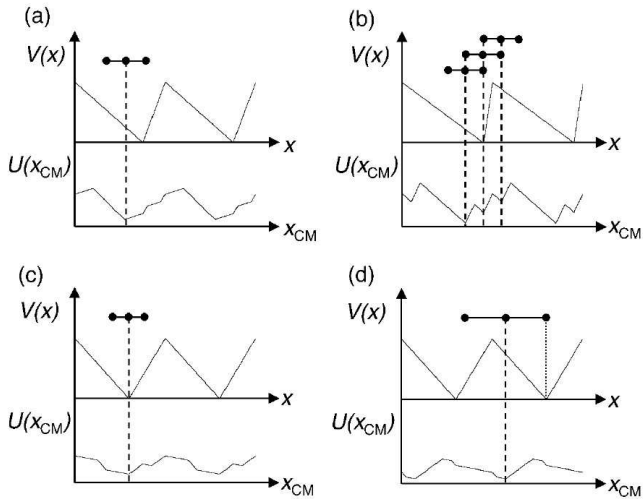}
  \caption{Trimers constrained to 1D motion ($kT \ll V_\mathrm{max}$) are shown at their localization positions in a ratchet potential $V(x)$ for the cases: (a) $\alpha<1/3$ and $d<\alpha L$; (b) $\alpha <1/3$ and $d>\alpha L$; (c) $\alpha > 1/3$ and $d<\alpha L$; and (d) $\alpha >1/3$ and $d>\alpha L$. Dashed line indicates how the center-of-mass position of the trimer corresponds to a local minimum of the effective potential $U(x_\mathrm{CM})$. }
  \label{fig:fig4}
\end{figure}

Regime I ($\alpha<1/3$, $d<\alpha L$): The force on one particle on the steep side of the potential is stronger than the force on two particles on the shallow side of the potential, and therefore the trimer will localize with one of the particles at the sawtooth potential minimum ($x=(1-\alpha) L$). Because the spacing between particles is not large enough for the trimer to straddle adjacent potential wells, the only center-of-mass localization position is $x_\mathrm{min}=(1-\alpha)L-d$ (see Fig.~\ref{fig:fig4}(a)), and therefore $\Delta x_{+} = L + d - x_\mathrm{min} = \alpha L + 2d$ and $\Delta x_{-}= x_\mathrm{min} - d = (1-\alpha)L-2d$. The condition $\Delta x_{+} = \Delta x_{-}$ yields

\begin{equation}
\label{acrit3}
\alpha_c=\frac{1}{2}-\frac{2d}{L}.
\end{equation}

Regime II ($\alpha<1/3$, $d>\alpha L$): Since the spacing between particles is greater than $\alpha L$, the trimer now has three possible localization positions (Fig.~\ref{fig:fig4}(b)). The asymmetry condition for reversal depends on the relative probability for the trimer to localize in each of these positions. For $\alpha<1/2$, the probability is greatest for the trimer to localize in the position furthest in the $-x$ direction ($x_\mathrm{min}=(1-\alpha)L-d$). Therefore, it is reasonable to expect $ \Delta x_+  >  \Delta x_- $ for some $\alpha < 1/2$, yielding a critical asymmetry $\alpha_c < 1/2$.

Regime III ($\alpha>1/3$, $d<\alpha L$): The force on two particles on the shallow side of the potential is now stronger than the force on one particle on the steep side of the potential, leading to the localization position: $x_\mathrm{min}=(1-\alpha) L$ (Fig.~\ref{fig:fig4}(c)). Now, $\Delta x_+ = L - x_\mathrm{min} = \alpha L$ and $\Delta x_- = x_\mathrm{min} = (1-\alpha )L$, which yields $\alpha_c = 1/2$, so the velocity is always positive in this regime. 

Regime IV ($\alpha > 1/3$, $d>\alpha L$): A trimer with $d<x_\mathrm{CM}<L+d$ at the end of \toff~will localize at $x_\mathrm{min}=(2-\alpha)L-d$, such that the leading bead is at the minimum of a $V(x)$ potential well (Fig.~\ref{fig:fig4}(d)). Now, $\Delta x_+ = L+d - x_\mathrm{min}= 2d - (1-\alpha ) L$ and $\Delta x_- = x_\mathrm{min} - d = (2-\alpha )L -2d$, yielding $\alpha_c = 3/2 - 2d/L$. Since we have $d/L>\alpha>1/3$, the reversal condition is never met for this regime, and the velocity is always positive.

\begin{figure}[htbp]
  \centering
  \includegraphics[width=\columnwidth]{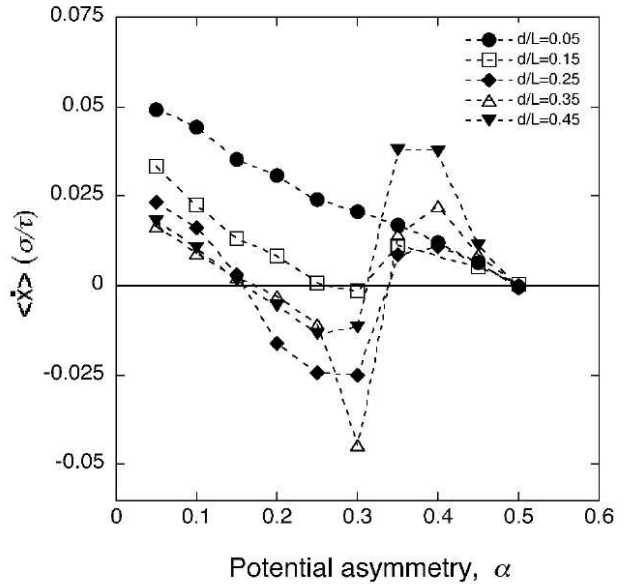}
  \caption{Brownian dynamics simulations. Average velocity as a function of $\alpha$ for trimers of length $d=0.05L$, $0.15L$, $0.25L$, $0.35L$ and $0.45L$, using $kT/V_\mathrm{max}=1/50$, $L=5$ and $\tau_{ON}=\tau_{OFF}=20$. The dashed-line interpolation between data points is included as a guide to the eye.}
  \label{fig:fig5}
\end{figure}

These predictions can be summarized as follows: The trimer has positive velocity for $1/3 < \alpha < 1/2$. When $\alpha<1/3$, $\alpha_c$ decreases with increasing $d/L$. When $d/L$ is large enough that $\alpha_c < 1/3$, then the trimer velocity reverses direction twice as $\alpha$ is increased from zero to 1/2:  first, from positive to negative velocity at $\alpha = \alpha_c$, and second, from negative back to positive velocity at $\alpha = 1/3$. However, if $d/L$ is small enough that $\alpha_c > 1/3$, then the trimer does not exhibit a velocity reversal with $\alpha$ for $\alpha < 1/2$, and the qualitative behavior of a monomer is recovered. Simulation results shown in Fig.~\ref{fig:fig5} confirm these analytical predictions. 

\begin{figure}[htbp]
  \centering
  \includegraphics[width=\columnwidth]{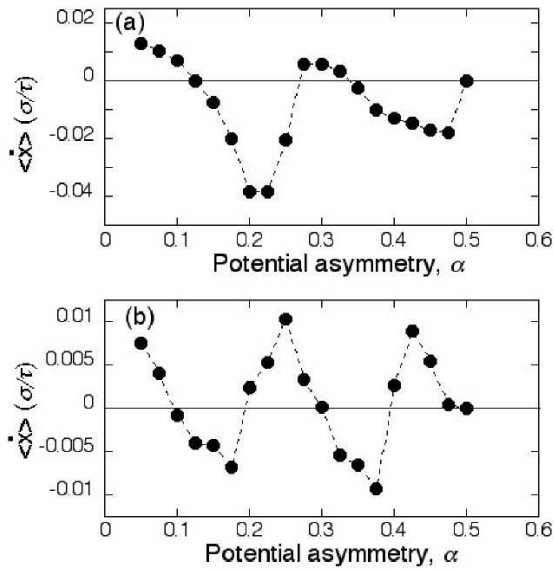}
  \caption{Brownian dynamics simulations. (a) Average velocity as a function of $\alpha$ is plotted for a chain of $N=4$ particles, using $d=0.2L$, $kT/V_\mathrm{max}=1/50$, $L=5$ and $\tau_{ON}=\tau_{OFF}=20$. The dashed-line interpolation between data points is included as a guide to the eye. (b) Average velocity as a function of $\alpha$ is plotted for a chain of $N=5$ particles, using the same parameters as in (a).}
  \label{fig:fignew}
\end{figure}

We have shown that the effective potential for a dimer in the presence of a `sawtooth' potential can have up to two minima in each period, and it is possible for the velocity to reverse once in the range $0<\alpha<1/2$.  Likewise, a trimer can have up to three minima in its effective potential and, for the right choice of parameters, the velocity reverses twice in the range $0<\alpha<1/2$. In Fig.~\ref{fig:fignew}, simulation results show that a chain with $N=4$ can undergo three velocity reversals in $0<\alpha <1/2$, and a chain with $N=5$ can reverse direction four times in this range. In general, if we consider a 1D chain of $N$ particles with $(N-1) d \leq L$, each period of the effective potential can have up to $N$ minima, and we expect ($N-1$) possible velocity reversals.

\section{Temperature dependent coupling effects\label{temp}}

Here, we discuss the temperature dependence of flashing ratchet velocity for mechanically coupled particles in 1D, compared with that of individual point particles. As we will show, the velocity of a dimer reverses direction twice as temperature increases, while a monomer in the same system does not reverse with temperature. First, we propose a reason for the ratchet velocity of a dimer to reverse direction with temperature, and then we present BD results demonstrating this behavior.

\begin{figure}[htbp]
  \centering
  \includegraphics[width=\columnwidth]{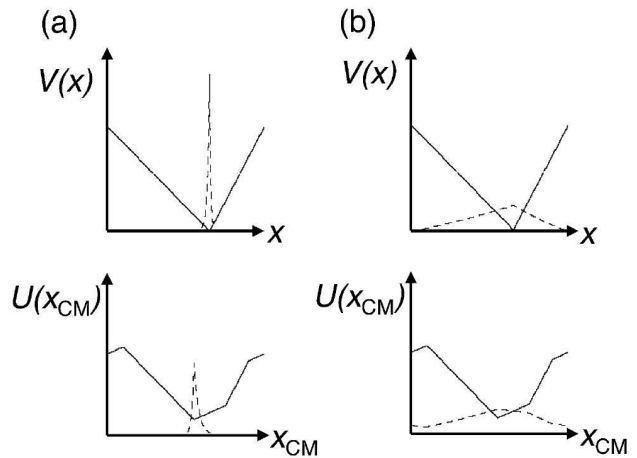}
  \caption{Upper panels: A ratchet potential, $V(x)$, with $\alpha=0.35$ is shown (solid line) along with the Boltzmann positional probability distribution (dashed line) for a monomer exposed to $V(x)$. Lower panels: $U(x_\mathrm{CM})$ is shown with the corresponding probability distribution for a dimer of length $d=0.2 L$. Probability distributions displayed for two choices of temperature: (a) $kT/V_\mathrm{max}=1/50$ and (b) $kT/V_\mathrm{max}=2$.}
  \label{fig:fig6}
\end{figure}

The velocity of an object in a flashing ratchet depends on the temperature of the surrounding heat bath in two ways: (1) Changing the temperature affects the magnitude of ratchet velocity by changing the diffusion constant, $D=kT/N \gamma_b$, of the chain. (2) An object in $V(x)$ localizes exactly at the minimum of $U(x_\mathrm{CM})$ only for the limit $kT \ll V_\mathrm{max}$. As temperature increases, the object's Boltzmann probability distribution broadens and the mean position shifts away from the effective potential minimum, $x_\mathrm{min}$, as illustrated in Fig.~\ref{fig:fig6}. For coupled particles, a period of $U(x_\mathrm{CM})$ has several regions of different relative slope, and the symmetry of the positional probability distribution shifts with increasing temperature. This can lead to a change in the direction of ratchet velocity. 

For a dimer with $x_\mathrm{min} < L/2$, where $x_\mathrm{CM}=0$ designates an absolute maximum of $U(x_\mathrm{CM})$, we expect ratchet velocity to be in the $-x$ direction in the limit ($kT \ll V_\mathrm{max}$), as discussed in section \ref{dimer}. However, as $T$ increases, we expect two velocity reversals: (A) From negative to positive velocity and (B) from positive back to negative velocity, for the following reasons:

Reversal (A): For a dimer exposed to $V(x)$, $U(x_\mathrm{CM})$ has a shallower slope immediately to the right ($+x$ direction) of the absolute minimum than to the left ($-x$ direction). Therefore, as $T$ increases from zero, the mean localization position increases, and a finite temperature can be chosen such that the dimer is more likely to be in the region ($L-\Delta < x_\mathrm{CM} < L$) during \ton~than in the region ($0 < x_\mathrm{CM} < \Delta$), where $\Delta$ is the dimer's average diffusion distance during  \toff, and is given by $\Delta = \sqrt{2D \mbox{\toff}}$. In this case, ratchet velocity is in the $+x$ direction, indicating a reversal from negative to positive velocity for increasing temperature.  

Reversal (B): Because a period of $U(x_\mathrm{CM})$ has several regions of different slope, continuing to increase the temperature can shift the probability distribution such that the dimer is more likely to be found in the region ($0< x_\mathrm{CM} <\Delta$) than in ($L-\Delta < x_\mathrm{CM} < L$), producing a second reversal from positive back to negative velocity. 

The physical reason for these reversals is that, at different temperatures, particles will sample regions of the potential that have different slopes. Because the effective potential for coupled particles is more complicated than the simple sawtooth potential $V(x)$, changing the temperature can change the symmetry of the center-of-mass probability distribution, producing a reversal in ratchet velocity.

 \begin{figure}[htbp]
  \centering
  \includegraphics[width=\columnwidth]{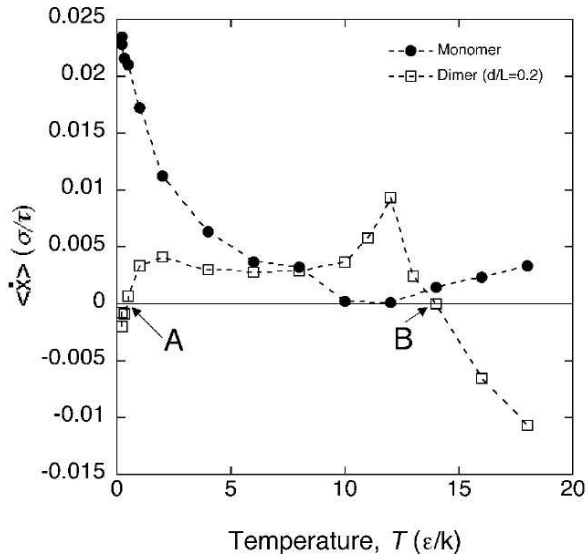}
  \caption{Brownian dynamics simulations. Average velocity as a function of temperature $T$ for a monomer and for a dimer of length $d=L/5$, using $\alpha=0.35$, $L=5.6$ and \ton~= \toff~= 20. The dashed-line interpolation between data points is included as a guide to the eye. Current reversals are labeled as points A and B (see text).}
  \label{fig:fig7}
\end{figure}
 
This prediction of two velocity reversals as a function of temperature for a dimer is confirmed by BD simulations, shown in Fig.~\ref{fig:fig7} along with the ratchet velocity of a single particle as a function of temperature. As temperature increases from $T=0$, the dimer undergoes a reversal from negative to positive velocity at a point labeled A. By further increasing the temperature, another velocity reversal is induced, at point B, from positive back to negative velocity. By contrast, the monomer always has positive velocity for $\alpha < 1/2$. 

 \section{Coupled particles in three dimensions\label{3Dmotors}}
 
Here, we examine the behavior of a dimer in a 3D system in the limit ($kT \ll V_\mathrm{max}$), in the same 1D external ratchet potential, $V(x)$, as in previous sections. When a dimer is allowed to rotate freely in 3D, the effective potential depends on the dimer's orientation with respect to the $x$-axis. The direction of velocity can be predicted based on the following insights: (1) When $V(x)$ is applied to a dimer with a random initial distribution, the dimer experiences both a linear force and a torque. In response to these forces, both particles in the dimer will localize at exactly the minimum of $V(x)$, and therefore the center-of-mass localization position is $x_\mathrm{min}=(1-\alpha)L$. (2) If \toff~is long enough that the dimer's orientation at the end of \toff~is uncorrelated with its orientation at the beginning, the average particle distribution of a dimer at the end of \toff~is a uniform shell with diameter $d$.
   
Based on these observations, it is possible to calculate the diffusion distances, $\Delta x_+$ and $\Delta x_-$, necessary for the dimer to localize in the adjacent well in the $+x$ and $-x$ direction, respectively, for the majority of possible dimer orientations at the beginning of \ton. To find these distances, we must determine the position $x_\mathrm{CM}$ for which the net force on the dimer, averaged over possible orientations, is zero in the presence of $V(x)$. It is straightforward to show that the net linear force on a uniform shell with diameter $d \le L$ exposed to $V(x)$ is zero if a fraction $\alpha$ of the diameter is to the left of a potential maximum. The probability distribution of a freely rotating dimer meets this requirement if it has $x_\mathrm{CM} = d/2-\alpha d$ or $x_\mathrm{CM} =L+d/2-\alpha d$. Thus, $ \Delta x_{+} = L+d/2-\alpha d - x_\mathrm{min} = \alpha L + d/2 - \alpha d$ and $ \Delta x_{-}  = x_\mathrm{min} - (d/2 - \alpha d) = (1-\alpha )L - d/2 + \alpha d$. The symmetry condition, $ \Delta x_{+}  =  \Delta x_{-} $, yields $\alpha_c = 1/2$, so there is no velocity reversal for $\alpha < 1/2$.

\begin{figure}[htbp]
  \centering
  \includegraphics[width=\columnwidth]{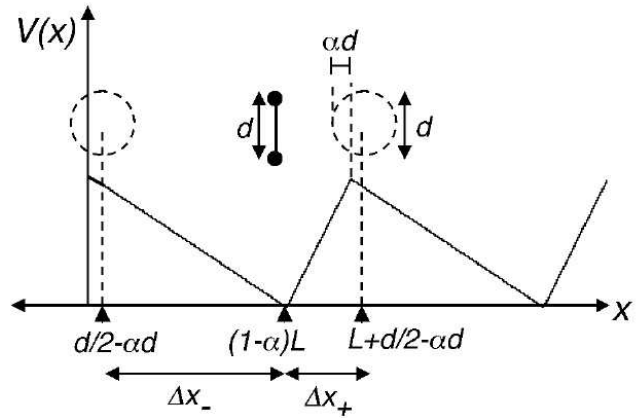}
  \caption{A freely rotating dimer ($d < \alpha L$) at $kT/V_\mathrm{max} \ll 1$ experiences on average no net linear force if $x_\mathrm{CM}= d/2 - \alpha d$ of $x_\mathrm{CM} = L+d/2- \alpha d$. The dashed-line circles indicate the positional probability distribution of a dimer centered on these $x_\mathrm{CM}$ positions at the end of \toff. The average diffusion distances necessary to localize in an adjacent well are labeled as  $\Delta x_{-} $ and $ \Delta x_{+}$. }
  \label{fig:fig8}
\end{figure}

\begin{figure}[htbp]
  \centering
  \includegraphics[width=\columnwidth]{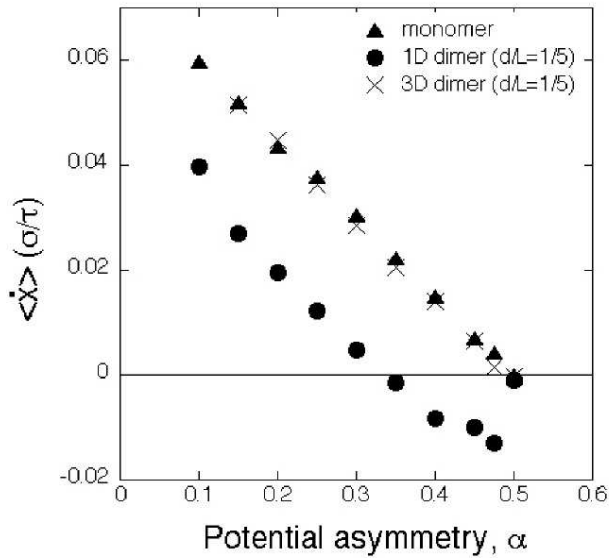}
  \caption{Brownian dynamics simulations. Center-of-mass velocity as a function of $\alpha$ for a freely-rotating dimer of length $d=0.2L$, using $kT/V_\mathrm{max}=1/50$, $L=5.6$ and \ton~$=$\toff~$=20$. For comparison, data points from Fig. 3a for a monomer and for a dimer confined to one-dimensional motion are also displayed.}
  \label{fig:fig9}
\end{figure}

This prediction is confirmed by BD simulations (Fig.~\ref{fig:fig9}). Results from Fig.~\ref{fig:fig3} are included in this plot to illustrate how the behavior of the freely-rotating dimer matches that of the monomer, in contrast to the velocity reversal that takes place for a dimer in 1D. We expect that the velocity reversal observed in the 1D system will also vanish for rigid chains of three or more particles in a 3D system, because the symmetry arguments used for the freely-rotating dimer will also apply to longer chains.

\section{Concluding remarks\label{discussion}}
 
We have shown that, in 1D, for $kT \ll V_\mathrm{max}$, a rigid chain of evenly spaced particles in a flashing ratchet reverses direction multiple times as a function of chain size or ratchet asymmetry. The physical reason is that coupled particles in a simple sawtooth potential are effectively equivalent to a single particle in a more complicated potential. In this sense, our results are related to the finding that a single particle in a ratchet driven by dichotomous force fluctuations can undergo multiple reversals when the external potential has multiple wells in each period \cite{kostur}.

Because a period of the effective potential of a dimer in $V(x)$ has multiple regions with different relative slope, the center-of-mass probability distribution can reverse symmetry as a function of temperature. The shifting of the probability distribution during \ton~changes the likelihood for the dimer to diffuse to an adjacent well in either direction during \toff. For this reason, a dimer can undergo multiple velocity reversals with temperature. Note that these reversals happen for a qualitatively different reason than the reversals described for coupled particles in the low temperature limit. In that limit, the direction of velocity is determined by the symmetry of the location of the confining minimum of the effective potential, relative to the absolute potential maximum. As a function of temperature, on the other hand, reversals occur because of the symmetry of the steepness of the confining potential. Due to the complexity of effective potentials for $N > 1$, the two reversal mechanisms are not always directly related.ÊÊ

While the velocity reversals observed in a 1D system do not occur for freely rotating rods in 3D, this does not rule out the possibility for reversal when the object's 3D rotation is partially confined. The ratchet transport of rigid rods confined to a tunnel with radius on the order of the rod length could provide a model for coupling effects that can be experimentally realized more readily than a true 1D system. For example, transport of single-stranded DNA fragments has been observed using a 2D array of asymmetrically spaced, micron-scale electrodes to create a flashing ratchet on a silicon chip \cite{bader}. A similar technique may be used to experimentally realize the present theory. For example, one may expose a polyelectrolyte that is confined in a quasi-1D nanochannel \cite{reisner, tegenfeldt1, tegenfeldt2} to a time-dependent, asymmetric electrostatic potential. Such a system could provide a method to separate particles of different sizes in opposite directions.

Our investigation of the flashing ratchet transport of mechanically coupled particles may also be relevant to the directed intracellular transport of linear molecular motors \cite{reimann3,astumian, astumian2}. It has been observed that structurally similar molecular motors can move in opposite directions \cite{henningsen,sablin,case,bloom,mcdonald,walker}. Binding between a motor head and its track is usually by multiple bonds, which one may choose to model by several, semi-rigidly interconnected ``beads" exposed to a binding potential. In dimeric motors, an unbound motor head is not free to rotate in 3D because of its attachment to the track by another motor head. The binding potential is thus quasi-1D, as discussed in the present paper. Our observation that objects with more internal structure have a larger number of velocity reversals may explain why, in complex motors, a relatively small change in structure can produce reversal of walking direction.

The current study deals with rigid rods which are shorter than a spatial period of the ratchet potential. In another recent study, we explored the role of flexibility in flashing ratchet transport by modeling a 3D Brownian motor based on a polyelectrolyte or polymer carrier with a radius of gyration on the order of several ratchet periods, demonstrating that flexibility can increase the speed and the stall force of the motor \cite{downton}. The unique collective effects observed in each of these cases suggest that it would be interesting to investigate an intermediate regime, such as self-avoiding, flexible chains in 3D with total contour length less than a ratchet period.

\section*{Acknowledgements}
\label{sec:ack}

We are grateful to Matthew Downton and Mike Plischke for helpful discussions. This work is supported by the National Science Foundation (PHY 0239764) to H.L., by NSF-IGERT and by NSF GK-12 (E.C.), and by a Discovery grant from NSERC of Canada (M.Z.). This research has been enabled by the use of the Bugaboo computing facility and the IRMACS center of Simon Fraser University, which are funded in part by the Canada Foundation for Innovation.

\end{document}